# Characterization of Blast Waveforms Produced by Different Driver Gasses in an Open-Ended Shock Tube Model

Evan L. Reeder*[1], Mei-Ling Liber*[2], Owen D. Traubert*[2], Christopher J. O'Connell[1], Ryan C. Turner[3], Matthew J. Robson[1]

[1]University of Cincinnati James L. Winkle College of Pharmacy, Division of Pharmaceutical Sciences, Cincinnati, OH
[2]University of Cincinnati College of Arts and Sciences, Department of Biological Sciences, Cincinnati, OH
[3]West Virginia University, Department of Neurosurgery, Morgantown, WV

*These authors contributed equally
Corresponding Author: Matthew J. Robson, matthew.robson@uc.edu

**Abstract:**
With the evolution of modern warfare and the increased use of improvised explosive devices (IEDs), there has been an increase in blast-induced traumatic brain injuries (bTBI) among military personnel and civilians. The increased prevalence of bTBI necessitates bTBI models that result in a properly scaled injury for the model organism being used. The primary laboratory model for bTBI is the shock tube, wherein a compressed gas ruptures a thin membrane, generating a shockwave. To generate a shock wave that is properly scaled from human to rodent subjects the shock wave must have a short duration and high peak overpressure while fitting a Friedlander waveform, the ideal representation of a blast wave. A large variety of factors have been experimentally characterized in attempts to create an ideal waveform, however we found current research on the gas composition being used to drive shock wave formation to be lacking. To better understand the effect the driver gas has on the waveform being produced, we utilized a previously established murine shock tube bTBI model in conjunction with several distinct driver gasses. In agreement with previous findings, helium produced a shock wave most closely fitting the Friedlander waveform in contrast to the plateau-like waveforms produced by some other gases. The peak pressure at the exit of the shock tube and 5 cm from the exit have a strong negative correlation with the density of the gas being used: helium the least dense gas used produces the highest peak overpressure. Density of the driver gas also exerts a strong positive effect on the duration of the shock wave, with helium producing the shortest duration wave. Due to its ability to produce a Friedlander waveform and produce a waveform following proper injury scaling guidelines, helium is an ideal gas for use in shock tube models for bTBI.

# Introduction

Modern warfare techniques, including the use of improvised explosive devices (IEDs), have increased the incidence of blast-induced injuries in both military and civilian populations across the globe. The most pronounced of these injuries includes blast-induced traumatic brain injury (bTBI). In the US military, an estimated 385,000 cases of TBI have occurred between the years 2000 and 2018,[1] with studies estimating nearly 20% incidence in ongoing military conflicts.[1-3] An estimated eightfold increase in blast injuries has occurred during this same timeframe in civilian populations,[4,5] extending the health relevance of these types of injuries. Those who suffer from TBIs experience increased rates of post-traumatic stress disorder (PTSD), depression, and suicide attempts.[6,7] These statistics outline the need for further study and the importance of models for bTBI that recapitulate the enduring clinical comorbidities associated with injury.

Military and improvised explosions consist of several different factors producing the potential for TBI. Sudden onsets of overpressure cause significant damage to the brain (primary blast injury), inducing alterations in intracranial pressure (ICP) and damage to the blood-brain barrier (BBB).[8] Other pressure-sensitive organs such as the ears and lungs are also at high risk of damage.[8] Due to the force of an explosion and strong blast winds, objects within the blast radius may be propelled outwards from the blast center.[9] This can include objects impacting the body and head causing blunt trauma or possibly penetrating the body or skull causing penetrating injury (secondary blast injury).[9,10] The force of the blast wave can also produce bodily movement wherein the whole body experiences rapid acceleration forces as it is projected away from the blast and subsequently a rapid deceleration as it collides with an object or the ground (tertiary blast injury).[10]

Blast waves are caused by a rapid expansion of gas during an explosion creating a blast overpressure. This overpressure decays and is followed by a negative pressure phase resulting from a relative vacuum.[11] Since the earliest work in modeling blast-induced brain injury, developing reproducible and clinically relevant models of wartime blast waves has been both a challenge and a major focus of the maturing field of bTBI.[12] Hooker's 1924 work using gun blasts and high explosives to cause bTBI, then called shock, was perhaps the first time the significance of the shape of blast waves in causing brain injury was recognized.[12] Waveforms that resulted in shock were characterized by a brief, sharp overpressure period with an immediate decay and subsequent negative phase, while blasts with longer duration overpressures, even with relatively higher overpressure magnitudes, caused numerous injuries but did not result in shock.[12] The waveform of the former variety of blast wave has been termed the Friedlander waveform (Figure 1).[13] In engineering terms, the Friedlander waveform depicts the ideal blast wave in an open field, and given its reproducibility, simplicity, and clinical relevance, it is the gold standard of blast waveforms generated by preclinical bTBI models.[10,11]

At present, there is not a sole standard preclinical model for bTBI; however, several models have been developed that account for the varied effects of bTBI that contribute to injury. Although the lack of standardization may seem problematic, bTBI is inherently heterogeneous in nature. Variability in models within parameters that are appropriate and clinically relevant (i.e. still scaled to the model organism employed and following the Friedlander waveform) may be a strength in studying and developing tools for the prevention and treatment of bTBI. Despite the lack of fine-detail standardization, nearly all the currently utilized preclinical models for bTBI use gas-driven shock tubes given the obvious complications of using explosives in a laboratory environment. Gas-driven shock tubes are designed to generate blast waves reminiscent of wartime explosions, i.e. a Friedlander waveform, through the use of a driver section separated from a driven section with a thin polymer diaphragm.[14-17] As gas fills the driver section, pressure builds causing the rupture of the diaphragm, and a pressure wave is generated propagating along the driven section. Although this general setup is fairly standardized, specific parameters of the shock tube such as length of driven/driver sections, diameter of the shock tube, position of subject, thickness of membrane utilized, and the gas used to drive the shock wave vary from model to model and may act to impact shock wave formation and propagation in ways that may ultimately impact the biological response imparted upon experimental subjects.[11, 16, 18-20]

Scaling the blast wave produced in a shock tube model for the specific animal subject being used ensures the model is representative of the blast conditions humans may experience, where body mass and surface area should be accounted for with regards to blast wave exposure.[10, 21] Since at least the 1960s and Bowen's work on scaling blasts for pulmonary injury, it has been widely recognized that the duration of the positive overpressure period, but not necessarily the value of the peak overpressure, needs to be scaled down for smaller model organisms to achieve accurate tissue injury for primary blast injury.[22] Subsequent work has confirmed that the principle of scaling blast duration is necessary to achieve accurate injury for bTBI caused by primary blast injury.[23-25] Further, longer duration blast waves have been shown to increase lethality and result in significant cell death in rodents.[26, 27] Thus, to ensure better injury scaling, efforts have been made across various blast models to decrease the blast duration and ensure the subjects aren't exposed to needlessly long blast overpressure durations (> 4 ms).[16] Other work has determined that due to differences in body area, the magnitude of the peak overpressure should be greater for model organisms with lower body area to achieve similar tertiary blast injuries to those seen in humans, with some suggesting a simple peak overpressure scaling factor of human body area over model organism body area.[10, 11] Thus, waveforms in shock tube bTBI models that seek to replicate blast injury in rodents must have a short temporal duration and have a high peak overpressure relative to wartime blasts.

A large variety of factors have been experimentally characterized in attempts to create an ideal waveform utilizing preclinical shock tube models. Factors such as position of the subject relative to the shock tube, energy reflections from the environment, and driver and driven section lengths have all been investigated thoroughly.[11, 16, 18-20] However, to our knowledge, few gases have been tested as driver gas to determine their impact on the waveform for TBI blast models. The field of engineering has, for decades, studied the effects of varying gas densities and/or molecular weights on shock wave formation,[28-30] However, a relatively limited number of studies have systematically studied this using laboratory shock tubes specifically designed for the induction of bTBI. The lack of standardization of the driver gas, as well as the failure for some shock tube studies to identify which driver gas was used necessitates a better understanding of the important role the driver gas plays in producing the shock waveform. Herein, we utilized a shock tube apparatus that has been characterized as a preclinical model for bTBI to determine the effects of varying gas compositions on waveform generated. We characterized the blast waves produced by gases of several different densities using this blast model for TBI. We find that different gas compositions drastically alter the generated waveforms stemming from this model, an effect that could have profound effects on scaling parameters and ultimately the biologic effects these models exert on vertebrate subjects.

## Methods

### Literature Search of Most Utilized Gasses

In order to more concretely establish what gas compositions are in use for driving shock tube models of bTBI, the Harzing Publish or Perish software (Harzing, A.W. (2007) Publish or Perish, available from https://harzing.com/resources/publish-or-perish) was used to search

Google scholar for the keywords "traumatic brain injury," "model," "blast," and "tube". The top 50 matching papers making use of shock tube blast models by citation count were identified and the gas composition was recorded. Papers were categorized as employing compressed air; helium; nitrogen; some combination of air, helium, and/or nitrogen; and not specified but citing a prior source or not specified at all. For the not specified but citing a prior source category, the gas used by the prior source was identified and recorded.

**Blast Model**

A previously described murine model for bTBI was utilized (Figure 2), to generate blast waves.[14, 15, 25] Briefly, a machined driver (9 cm total length, 7 cm internal length) section containing a 6-degree angle and driven (15 cm) section made from high-tensile steel compose the shock tube.[31] The driver and driven section were separated by clear 76.2 μm thick mylar membrane for all data contained within the current manuscript. All high-pressure gas cylinders utilized were connected to a high flow gas regulator (Harris) with output pressure standardized to 2585 kPa (375 psi). Gas cylinders are connected to the driver section through high-pressure quick-connect hoses with a dead man's lever ensuring rapid cessation of gas release.

**Driver Gasses**

Compressed gases of varying densities: medical-grade atmospheric air (Atmospheric Air; 21% $O_2$, 79% $N_2$, Airgas #USP200), Helium (He; Airgas #HE300), Argon (Ar; Wright Brothers, Inc.), Nitrogen ($N_2$; Wright Brothers, Inc.) Carbon Dioxide ($CO_2$; Wright Brothers, Inc.) were commercially certified and obtained (Table 1). Each compressed gas cylinder was connected to the Blast Model and used to produce several blast waves for subsequent analysis.

**Data Acquisition and Sensor Apparatus**

Three high frequency ICP® pressure sensors, model 102A05 (PCB Piezoelectrics, Sensitivity 7.3 mV/kPa), were placed at the exit of the driven section 120° apart. An additional sensor was placed 5 cm from the end of the driven section in order to record the incident blast wave. Dynamic pressure measurements were initiated immediately prior blast implementation and recorded at a sampling frequency of 500,000 frames per second using a sensor signal conditioner (PCB Piezoelectrics, 482C series 4-channel signal conditioner) and data acquisition board (National Instruments, Labview version 12.0). Pressure readings were captured in pounds per square inch (psi) and converted to kilopascal (kPa). Each blast condition for each gas was replicated (Atmospheric Air n=5, $CO_2$ n=10, Ar n=8, $N_2$ n=10, He n=7) for a total of 40 blast waves analyzed.

**Blast Waveform Analysis**

To create the wave plots, pressure sensor data for each gas were stacked into a 3D array in Matlab version 2020b and averaged against time in milliseconds. The incident pressure sensor (see Figure 2) was used to calculate the peak incident pressure, positive impulse, and the phase duration for each gas. The blast profiles were offset to ensure that they could be superimposed on one another given a start time of 2 ms. Peak incident pressure was determined by finding the maximum pressure value for each trial and calculating an average for each gas. Positive impulse was determined by creating a gridded interpolant for each profile and integrating over the positive phase portion of the Friedlander curve, using the time corresponding to the peak incident pressure and the time at which the pressure transitioned to the negative phase as endpoints. Positive phase duration was computed by the same methodology used to find the positive impulse and was the time from the onset of peak overpressure to the point where the line crossed 0 kPa and began the negative phase.

**Statistical Analyses**

All statistical analyses were conducted in GraphPad Prism 8. Peak pressure, positive impulse, and positive phase duration between the five gases were compared using a one-way analysis of variance (ANOVA) followed by Dunnett's post hoc tests comparing each gas to atmospheric air as the control. Density correlation was analyzed using a linear regression. Linear regression lines are shown with 95% confidence intervals. For all statistical analyses, $P ≤ 0.05$ was considered statistically significant. All results are presented as Mean ± SEM.

## Results

### Literature Search of Driver Gas Composition

A literature search of the top 50 most cited publications utilizing gas-driven shock tube bTBI models was conducted and the identity of the driver gas was recorded. Analysis revealed 52% (n=26) used compressed air, 32% (n=16) used helium, 4% (n=2) used nitrogen, 4% (n=2) used multiple gasses, and 8% (n=4) did not specify the identity of the gas used (Figure 3).

### Peak Pressures

Peak pressure recordings were collected at both the exit of the driven section and 5 cm away from the exit, measuring exit pressure and the incident wave respectively. Average peak exit pressure and peak incident pressure for each gas was plotted against its density. He, the least dense gas used in these studies, produced a blast wave with the highest peak pressure at the exit of the driven section whereas $CO_2$, the densest gas used, produced the smallest peak pressure at the exit of the driven section (Figure 4A; Atmospheric Air=222.73±6.99 kPa, He=376.55±15.45 kPa, $N_2$=218.79±4.10 kPa, Ar=185.59±6.72 kPa, $CO_2$=179.71±3.52 kPa; one-way ANOVA $P<0.0001$, He vs. Atmospheric air $P<0.0001$, $N_2$ vs. Atmospheric Air $P=0.9886$, Ar vs. Atmospheric Air $P<0.5$, $CO_2$ vs. Atmospheric Air $P<0.01$). Peak pressure at the exit of the shock tube displayed a strong dependence on the density of gas used (Figure 4B; He (0.1664g/L)=376.55 kPa, Atmospheric Air (1.204g/L)=222.73 kPa, $N_2$ (1.165g/L)=218.79 kPa, Ar (1.661g/L)=185.59 kPa, $CO_2$ (1.842g/L)=179.71 kPa; $R^2=0.95$).

Similar results were observed in the peak incident pressure with He producing the highest peak incident pressure and $CO_2$ producing the smallest peak incident pressure (Figure 4C; Atmospheric Air=707.95±46.05 kPa, He=1162.16±89.30 kPa, $N_2$=631.58±18.15 kPa, Ar=556.60±17.22 kPa, $CO_2$=388.98±16.74 kPa; one-way ANOVA $P<0.0001$, He vs. Atmospheric air $P<0.0001$, $N_2$ vs. Atmospheric Air $P=0.4972$, Ar vs. Atmospheric Air $P=0.0688$, $CO_2$ vs. Atmospheric Air $P<0.0001$). There also exists a strong positive linear correlation between peak incident pressure and the density of the gas used (Figure 4D; He (0.1664g/L)=1162.16, Atmospheric Air (1.204g/L)=707.95, $N_2$ (1.165g/L)=631.58, Ar (1.661g/L)=556.60, $CO_2$ (1.842g/L)=388.98; $R^2=0.96$).

### Shockwave Duration

The duration of the incident pressure wave was determined and plotted for each of the driver gasses. The driver gas used displayed a great deal of influence over the duration of the positive pressure phase of the blast wave. He gas produced the shortest blast wave duration, 1.03 ms and $CO_2$ produced a blast wave with the longest duration, 2.29 ms (Figure 5A; Atmospheric Air=1.57±0.14 ms, He=1.03±0.03 ms, $N_2$=1.56±0.09, Ar=1.85±0.49 ms, $CO_2$=2.29±0.07 ms; one-way ANOVA $P<0.0001$, He vs. Atmospheric air $P<0.05$, $N_2$ vs. Atmospheric Air $P>0.9999$, Ar vs. Atmospheric Air $P=0.2803$, $CO_2$ vs. Atmospheric Air $P<0.001$). There is a strong negative correlation between the density of the driver gas and the duration of the wave produced, with lower density gasses producing shorter duration blast waves (Figure 5B;

He (0.1664g/L)=1.03 ms, Atmospheric Air (1.204g/L)=1.57 ms, $N_2$ (1.165g/L)=1.56 ms, Ar (1.661g/L)=1.85 ms, $CO_2$ (1.842g/L)=2.29 ms; $R^2$=0.91).

**Positive Impulse**

Positive impulse is the area under the positive phase on the pressure versus time graph of the incident pressure waves. This value was calculated for each of the gasses used. He gas produced the lowest positive impulse whereas $CO_2$ produced the largest positive impulse (Figure 6A; Atmospheric Air=112.63±3.32 kPa*ms, He=69.01±1.50 kPa*ms, $N_2$=110.52±7.00 kPa*ms, Ar=106.25±0.49 kPa*ms, $CO_2$=164.78±5.89 kPa*ms; one-way ANOVA $P<0.0001$, He vs. Atmospheric air $P<0.0001$, $N_2$ vs. Atmospheric Air $P=0.9961$, Ar vs. Atmospheric Air $P=0.8512$, $CO_2$ vs. Atmospheric Air $P<0.0001$). The impulse produced by atmospheric air, Ar, and $N_2$ were all relatively similar to one another. There is a positive linear correlation between the positive impulse produced and the density of the gas (Figure 6B; He (0.1664g/L)=69.01 kPa*ms, Atmospheric Air (1.204g/L)=112.63 kPa*ms, $N_2$ (1.165g/L)=110.52 kPa*ms, Ar (1.661g/L)=106.25 kPa*ms, $CO_2$ (1.842g/L)=164.78 kPa*ms; $R^2$=0.72).

**Discussion**

Gas-driven shock tube models are the most commonly used method to reproduce blast overpressures for preclinical modeling of neurotrauma incurred by military personnel. Due to the high rate of bTBI among the military, reliable, reproducible models are crucial to the basic understanding of how blast exposure impacts biological substrates and the brain. A variety of key factors affect blast waves produced by shock tube models commonly utilized in preclinical studies and many of these factors have been extensively investigated and modeled previously.[16, 18, 19] However, we have found the available data on the composition of the compressed gas used to drive the generation of blast waves within these models to be lacking. The gas driving the blast wave is a key element affecting the waveform produced and thus a more thorough comparison including more driver gases was needed to fill this gap. We systematically characterized and analyzed blast waves generated in a single model, driven by several gasses: helium, atmospheric air, nitrogen, argon, and carbon dioxide to overtly study the effect that the driver gas has on the blast wave produced and its parameters. We identified some of the key characteristics of the blast wave: peak overpressure, duration, and impulse across different driver gases as well as adherence to the ideal Friedlander waveform.

Previous studies have reported that He gas produces a waveform that undergoes a rapid exponential decay consistent with what is expected of a Friedlander waveform, while some gasses such as $N_2$ have a flattop or plateau wave.[20] The overlay of the blast waves produced by He and $CO_2$ demonstrate the differing pressure decay waveform between these two driver gasses (Figure 7f). Immediately after the peak pressure occurs, the wave produced by He undergoes a rapid decay resulting in a short duration of the positive pressure phase. Conversely, the blast wave produced by $CO_2$ has a plateau period before the exponential decay occurs. This plateau effect can also be seen in the graphs of atmospheric air, $N_2$, and Ar (Figure 7a, c, and d). Given the established need in the field for an idealized, standardized blast curve, He gas is thus foremost among driver gas candidates for its ability to match the Friedlander waveform without any plateau artifacts.

Proper injury scaling is essential to ensure the translatability of pre-clinical models for TBI to humans. For bTBI this scaling comes from aspects of the blast wave such as peak pressure, duration of the wave, and impulse.[10, 23] Based on brain mass duration scaling, depending on the rodent species used, blast waves produced by rodent bTBI models should be from 0.04 to 0.07 times as long as wartime blasts.[24, 32] Modeling of a mortar grenade indicates a positive overpressure phase duration of around 15 ms 10 meters from the charge, and experiments using real high explosive charges of significantly larger size found positive overpressure duration of 14.5 and 18.2 ms at 30 and 24 meters from the charge respectively.[33] Shridharani et al. suggest 1.3 to 6.9 ms are realistic for similar survivable blast.[34] Based on our approximate brain mass scaling factor, rodent bTBI models seeking to replicate these kinds of wartime blasts should have positive overpressure durations in the approximate range of 0.05-1.3 ms depending on the size of the rodent species used and the blast modeled. Based on their estimation of human blast exposure of 1-10 ms, Needham et al. Suggest 0.1-1 ms is suitable in mice, while Wood et al. Suggest 1 ms. The overpressures produced by our model are all on a shorter timeframe (1-3 ms).[24, 35] He gas displays a duration of 1.03 ms while still producing a large peak overpressure. The other gasses used produced much lower peak incident pressures while at the same time extending the duration of the blast wave. These longer durations do not accurately scale to cause commensurate injuries between rodents and humans and may create excessive injury given experimental results indicating longer peak overpressures lead to unrealistically severe injury in rodents.

Based on area ratio acceleration scaling parameters, peak overpressure for rodent models should be around 65-100 times greater than wartime blasts depending on the rodent species.[10, 11, 36] In wartime blasts, 1 psi peak overpressure is enough to knock a soldier to the ground, 5 psi peak overpressure is enough to rupture the tympanic membrane, and 8 psi peak overpressure is enough to knock over a railcar.[37] Given tertiary blast injury typically requires interaction with the environment, it is difficult to establish a concrete range for injury. However, it seems reasonable to suggest that pressures greater than that required to knock over a soldier would be required for tertiary blast TBI. Modeling studies have determined blast overpressures ranging from 15.5-107.5 psi to be representative of survivable wartime blasts.[38] Therefore, we suggest 130 psi (896 kPa) or greater is necessary to model tertiary blast injury in rodents based on applicable scaling factors of body mass and surface area.[36] He gas has the highest peak overpressure of 1162 kPa at incident, much higher than the closest second, atmospheric air at 708 kPa; gasses other than helium have peak incident overpressures that may result in unrealistically minor tertiary blast injury. Our studies, in concordance with others,[16, 17, 20] have established that He, as the gas candidate with the shortest blast duration and coincident highest peak pressure, is thus foremost in its ability to scale bTBI to a rodent model. While similar scaling of the resulting waveform is possible by established methods (e.g. modifying tube length), the use of helium gas in our shock tube accomplished this without any significant modification of the apparatus.

Blast models for TBI have been developed with the aim of producing a Friedlander waveform to model the blast waves produced by IED explosions. As we observed, while He gas produces blast waves similar to the ideal Friedlander waveform, other gases produce blast waves with a plateau effect prior to their exponential decay. These results are concordant with the limited studies conducted by others within the field using other models for bTBI.[20] He gas also had the highest peak overpressure and shortest duration blast wave of all candidate gases, which is necessary for proper injury scaling for use in rodent subjects.[10, 22-24] Our data points to He being an ideal gas for modeling bTBI rodent subjects, in agreement with others in the field, whose published data on their model waveforms appears to resemble our own.[16, 17, 20] Due to its ability to produce a Friedlander waveform and produce a waveform following proper injury scaling guidelines, He is an ideal gas for use in shock tube models for bTBI. Many rodent bTBI models utilize compressed air to drive the blast wave, producing long overpressure durations, and causing the subject to experience a blast wave many times greater than the human equivalent.[10] As some others have previously suggested,[39] we concur that models that currently utilize compressed air or a gas other than He may be able to utilize He within their models to better fit the Friedlander waveform and produce a gas wave with a higher peak overpressure and shorter duration.

**Acknowledgments:**

This work was supported by the University of Cincinnati Foundation, University of Cincinnati Office of Research, a Brain and Behavior Research Foundation NARSAD Young Investigator Award (#25230, MJR), and a PhRMA Foundation Research Starter Grant (MJR).

Figure 2 was created using BioRender, biorender.com
Figure 3 was created using Harzing, A.W. (2007) Publish or Perish, available from https://harzing.com/resources/publish-or-perish

# References

1. Swanson TM, Isaacson BM, Cyborski CM, French LM, Tsao JW and Pasquina PF. Traumatic Brain Injury Incidence, Clinical Overview, and Policies in the US Military Health System Since 2000. *Public Health Rep*. 2017;132:251-259.
2. T. T and LH. J. Invisible wounds of war: psychological and cognitive injuries, their consequences, and services to assist recover. . *RAND Corporation Monograph Series* 2008.
3. McCabe JT and Tucker LB. Sex as a Biological Variable in Preclinical Modeling of Blast-Related Traumatic Brain Injury. *Front Neurol*. 2020;11:541050.
4. Mathews ZR and Koyfman A. Blast Injuries. *J Emerg Med*. 2015;49:573-87.
5. Wolf SJ, Bebarta VS, Bonnett CJ, Pons PT and Cantrill SV. Blast injuries. *Lancet*. 2009;374:405-15.
6. Wilks CR, Morland LA, Dillon KH, Mackintosh MA, Blakey SM, Wagner HR and Elbogen EB. Anger, social support, and suicide risk in U.S. military veterans. *J Psychiatr Res*. 2019;109:139-144.
7. Fann J and Hart T. Depression after traumatic brain injury. *Arch Phys Med Rehabil*. 2013;94:801-2.
8. Kawoos U, Gu M, Lankasky J, McCarron RM and Chavko M. Effects of Exposure to Blast Overpressure on Intracranial Pressure and Blood-Brain Barrier Permeability in a Rat Model. *PloS one*. 2016;11:e0167510-e0167510.
9. Phipps H, Mondello S, Wilson A, Dittmer T, Rohde NN, Schroeder PJ, Nichols J, McGirt C, Hoffman J, Tanksley K, Chohan M, Heiderman A, Abou Abbass H, Kobeissy F and Hinds S. Characteristics and Impact of U.S. Military Blast-Related Mild Traumatic Brain Injury: A Systematic Review. *Frontiers in neurology*. 2020;11:559318-559318.
10. Bass CR, Panzer MB, Rafaels KA, Wood G, Shridharani J and Capehart B. Brain Injuries from Blast. *Annals of Biomedical Engineering*. 2012;40:185-202.
11. Cernak I. Blast Injuries and Blast-Induced Neurotrauma: Overview of Pathophysiology and Experimental Knowledge Models and Findings. In: F. H. Kobeissy, ed. *Brain Neurotrauma: Molecular, Neuropsychological, and Rehabilitation Aspects* Boca Raton (FL): CRC Press/Taylor & Francis © 2015 by Taylor & Francis Group, LLC.; 2015.
12. Hooker DR. PHYSIOLOGICAL EFFECTS OF AIR CONCUSSION. *American Journal of Physiology-Legacy Content*. 1924;67:219-274.
13. Friedlander FG and Taylor GI. The diffraction of sound pulses I. Diffraction by a semi-infinite plane. *Proceedings of the Royal Society of London Series A Mathematical and Physical Sciences*. 1946;186:322-344.
14. Logsdon AF, Lucke-Wold BP, Turner RC, Collins SM, Reeder EL, Huber JD, Rosen CL, Robson MJ and Plattner F. Low-intensity Blast Wave Model for Preclinical Assessment of Closed-head Mild Traumatic Brain Injury in Rodents. *J Vis Exp*. 2020.
15. Logsdon AF, Lucke-Wold BP, Turner RC, Li X, Adkins CE, Mohammad AS, Huber JD, Rosen CL and Lockman PR. A Mouse Model of Focal Vascular Injury Induces Astrocyte Reactivity, Tau Oligomers, and Aberrant Behavior. *Arch Neurosci*. 2017;4.
16. Panzer MB, Matthews KA, Yu AW, Morrison B, 3rd, Meaney DF and Bass CR. A Multiscale Approach to Blast Neurotrauma Modeling: Part I - Development of Novel Test Devices for in vivo and in vitro Blast Injury Models. *Frontiers in neurology*. 2012;3:46-46.
17. Reneer DV, Hisel RD, Hoffman JM, Kryscio RJ, Lusk BT and Geddes JW. A multi-mode shock tube for investigation of blast-induced traumatic brain injury. *J Neurotrauma*. 2011;28:95-104.
18. Bandak FA, Ling G, Bandak A and De Lanerolle NC. Injury biomechanics, neuropathology, and simplified physics of explosive blast and impact mild traumatic brain injury. *Handb Clin Neurol*. 2015;127:89-104.
19. Kumar R and Nedungadi A. Using Gas-Driven Shock Tubes to Produce Blast Wave Signatures. *Frontiers in neurology*. 2020;11.

20. Sundaramurthy A and Chandra N. A parametric approach to shape field-relevant blast wave profiles in compressed-gas-driven shock tube. *Frontiers in neurology*. 2014;5:253.
21. Jean A, Nyein MK, Zheng JQ, Moore DF, Joannopoulos JD and Radovitzky R. An animal-to-human scaling law for blast-induced traumatic brain injury risk assessment. *Proceedings of the National Academy of Sciences*. 2014;111:15310-15315.
22. Bowen IG, Fletcher ER and Richmond DR. ESTIMATE OF MAN'S TOLERANCE TO THE DIRECT EFFECTS OF AIR BLAST. 1968.
23. Panzer MB, Wood GW and Bass CR. Scaling in neurotrauma: How do we apply animal experiments to people? *Experimental Neurology*. 2014;261:120-126.
24. Wood GW, Panzer MB, Yu AW, Rafaels KA, Matthews KA and Bass CRD. Scaling in blast neurotrauma. 2012.
25. Turner RC, Naser ZJ, Logsdon AF, DiPasquale KH, Jackson GJ, Robson MJ, Gettens RT, Matsumoto RR, Huber JD and Rosen CL. Modeling clinically relevant blast parameters based on scaling principles produces functional & histological deficits in rats. *Exp Neurol*. 2013;248:520-9.
26. Cernak I, Merkle AC, Koliatsos VE, Bilik JM, Luong QT, Mahota TM, Xu L, Slack N, Windle D and Ahmed FA. The pathobiology of blast injuries and blast-induced neurotrauma as identified using a new experimental model of injury in mice. *Neurobiology of Disease*. 2011;41:538-551.
27. Svetlov SI, Prima V, Kirk DR, Gutierrez H, Curley KC, Hayes RL and Wang KKW. Morphologic and Biochemical Characterization of Brain Injury in a Model of Controlled Blast Overpressure Exposure. *Journal of Trauma and Acute Care Surgery*. 2010;69:795-804.
28. Mozzhilkin VV. Motion of a shock wave through a gas of variable density. *Fluid Dynamics*. 1970;5:770-774.
29. Guo L and Wang X. Effect of molecular weight and density of ambient gas on shock wave in laser-induced surface nanostructuring. *Journal of Physics D: Applied Physics*. 2008;42:015307.
30. Freiwald DA. Approximate Blast Wave Theory and Experimental Data for Shock Trajectories in Linear Explosive-Driven Shock Tubes. *Journal of Applied Physics*. 1972;43:2224-2226.
31. Logsdon AF, Lucke-Wold BP, Turner RC, Collins SM, Reeder EL, Huber JD, Rosen CL, Robson MJ and Plattner F. Low-intensity blast wave injury model for preclinical assessment of closed-head mild traumatic brain injury in rodents. *Journal of Visualized Experiments*. 2020;Accepted in press.
32. Herculano-Houzel S. The human brain in numbers: a linearly scaled-up primate brain. *Frontiers in Human Neuroscience*. 2009;3.
33. Burghard Hillig AR, Hendrik Rothe. Numerical Investigation of the Attenuation of Shock Waves by Simulating a Second, Transient Medium to Protect Vehicles Against Blasts. *International Journal On Advances in Systems and Measurements*. 2018;11.
34. Pun PBL, Kan EM, Salim A, Li Z, Ng KC, Moochhala SM, Ling E-A, Tan MH and Lu J. Low level primary blast injury in rodent brain. *Frontiers in neurology*. 2011;2:19-19.
35. Needham EJ, Helmy A, Zanier ER, Jones JL, Coles AJ and Menon DK. The immunological response to traumatic brain injury. *Journal of neuroimmunology*. 2019;332:112-125.
36. Nair AB and Jacob S. A simple practice guide for dose conversion between animals and human. *Journal of basic and clinical pharmacy*. 2016;7:27-31.
37. Marklund N. Blast-Induced Brain Injury. In: T. Sundstrøm, P.-O. Grände, T. Luoto, C. Rosenlund, J. Undén and K. G. Wester, eds. *Management of Severe Traumatic Brain Injury: Evidence, Tricks, and Pitfalls* Cham: Springer International Publishing; 2020: 109-113.
38. Shridharani J, Wood G, Panzer M, Capehart B, Nyein M, Radovitzky R and Bass C. Porcine Head Response to Blast. *Frontiers in neurology*. 2012;3.
39. Chandra N, Ganpule S, Kleinschmit NN, Feng R, Holmberg AD, Sundaramurthy A, Selvan V and Alai A. Evolution of blast wave profiles in simulated air blasts: experiment and computational modeling. *Shock Waves*. 2012;22:403-415.

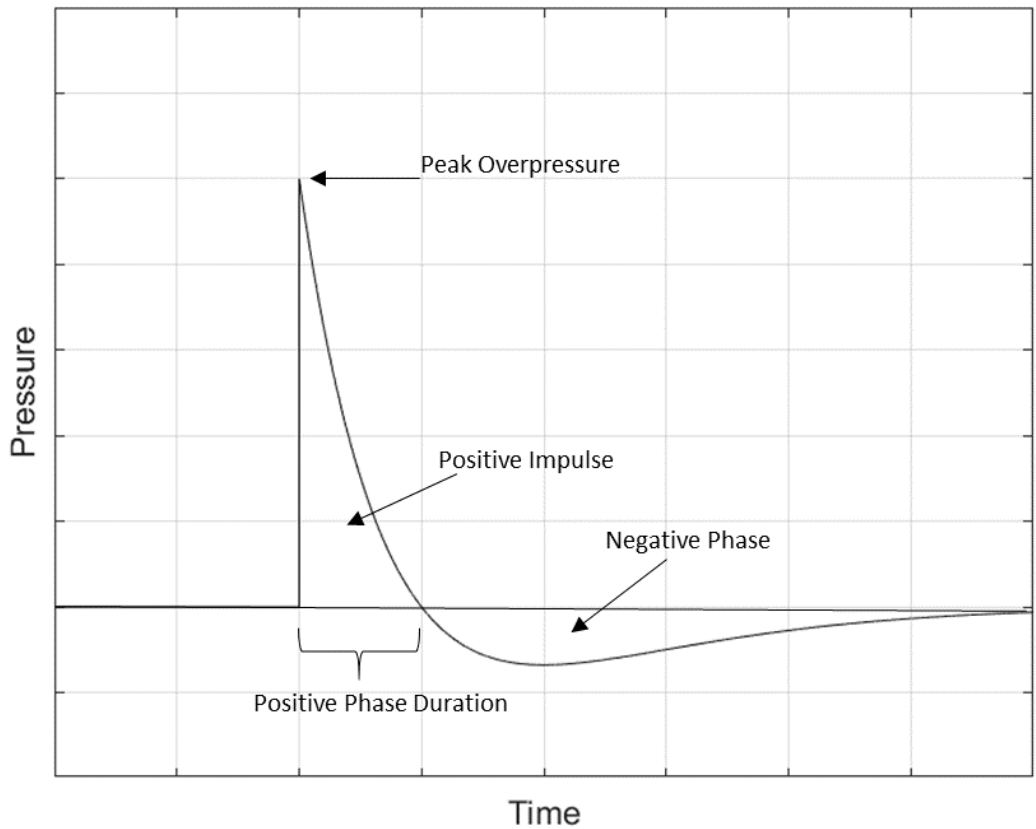

**Figure 1.** Friedlander waveform. The ideal blast waveform can be characterized by and is termed the Friedlander waveform. This consists of a rapid peak overpressure, and exponential decay, followed by a relative negative pressure phase. The area under the blast curve is termed the positive impulse.

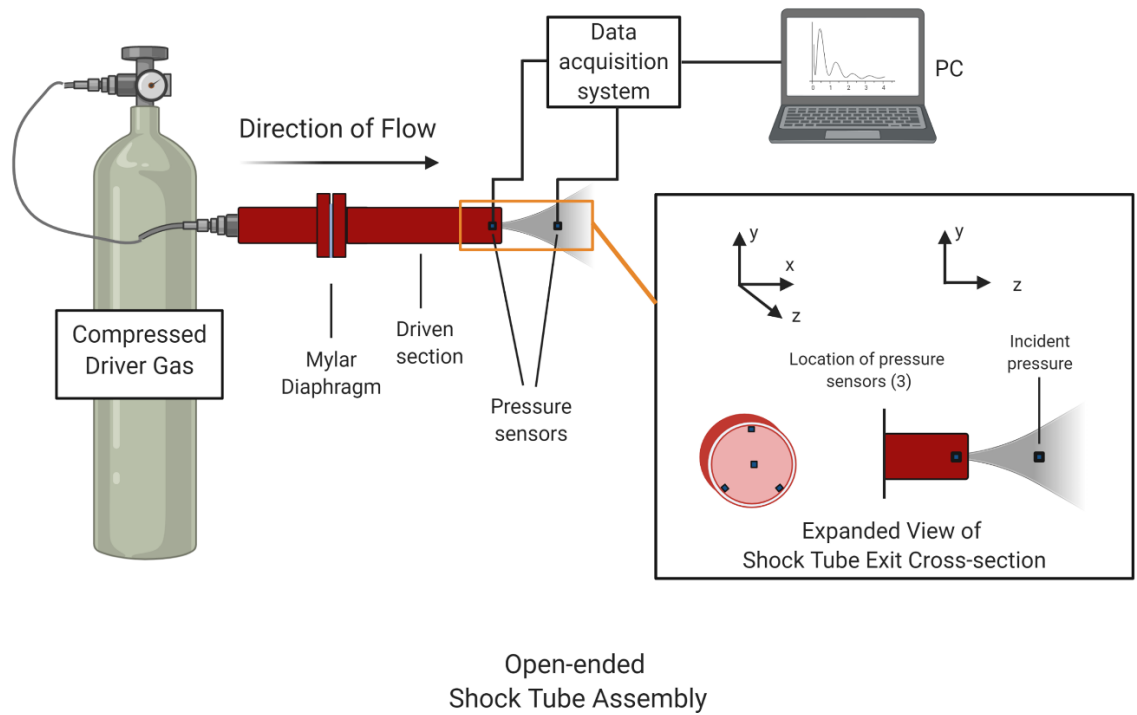

**Figure 2.** Diagram of blast model setup and sensor locations. Compressed gas (atmospheric air, helium, nitrogen, argon, carbon dioxide) pressurizes a Mylar membrane causing it to rupture and a pressure wave to propagate through the shock tube. Pressure sensors are placed at the exit of the driven section and 5 cm from the driven section to record the incident wave.

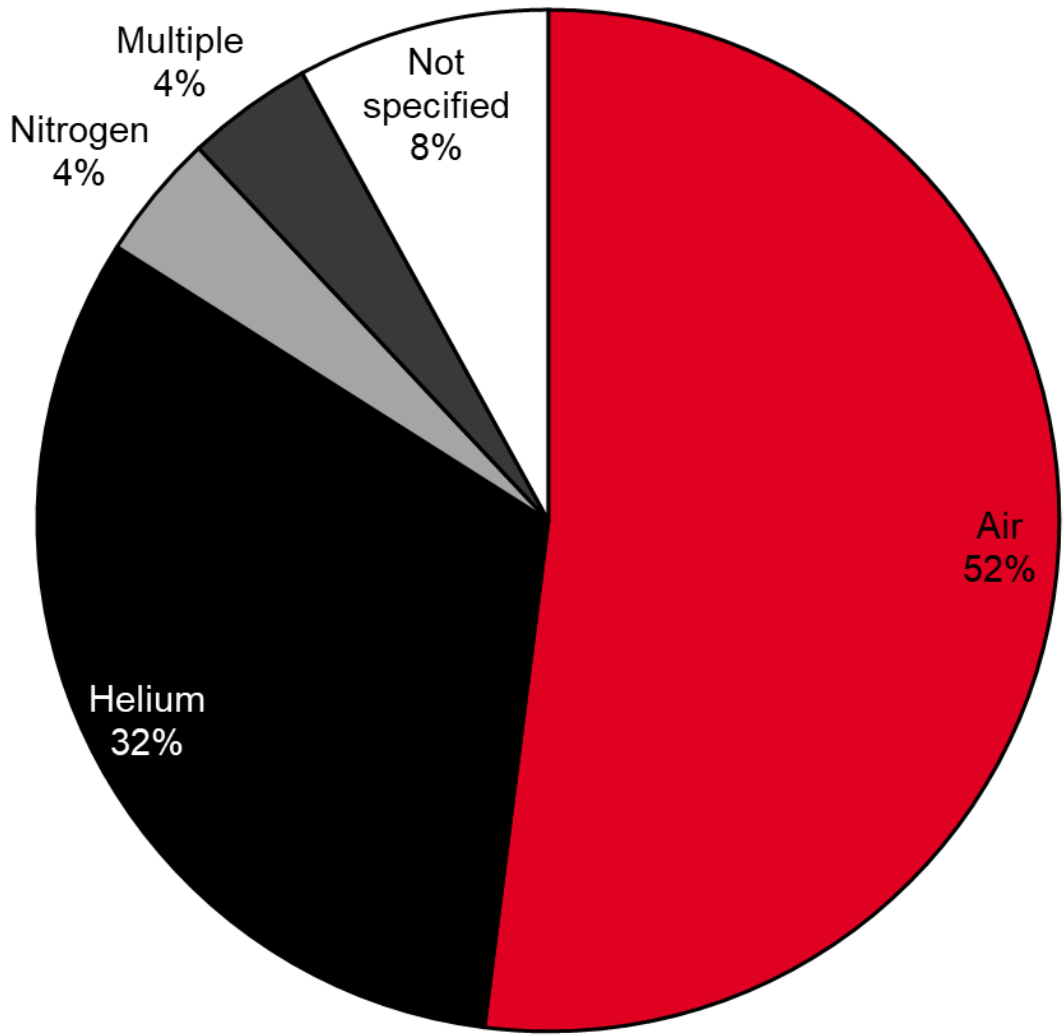

**Figure 3.** Literature search of driver gas use. A literature search of the top 50 most cited gas-driven shock tube bTBI models, revealed 52% (26) used compressed air, 32% (16) used helium, 4% (2) used nitrogen, 4% (2) used multiple gasses, and 8% (4) did not specify the gas used.

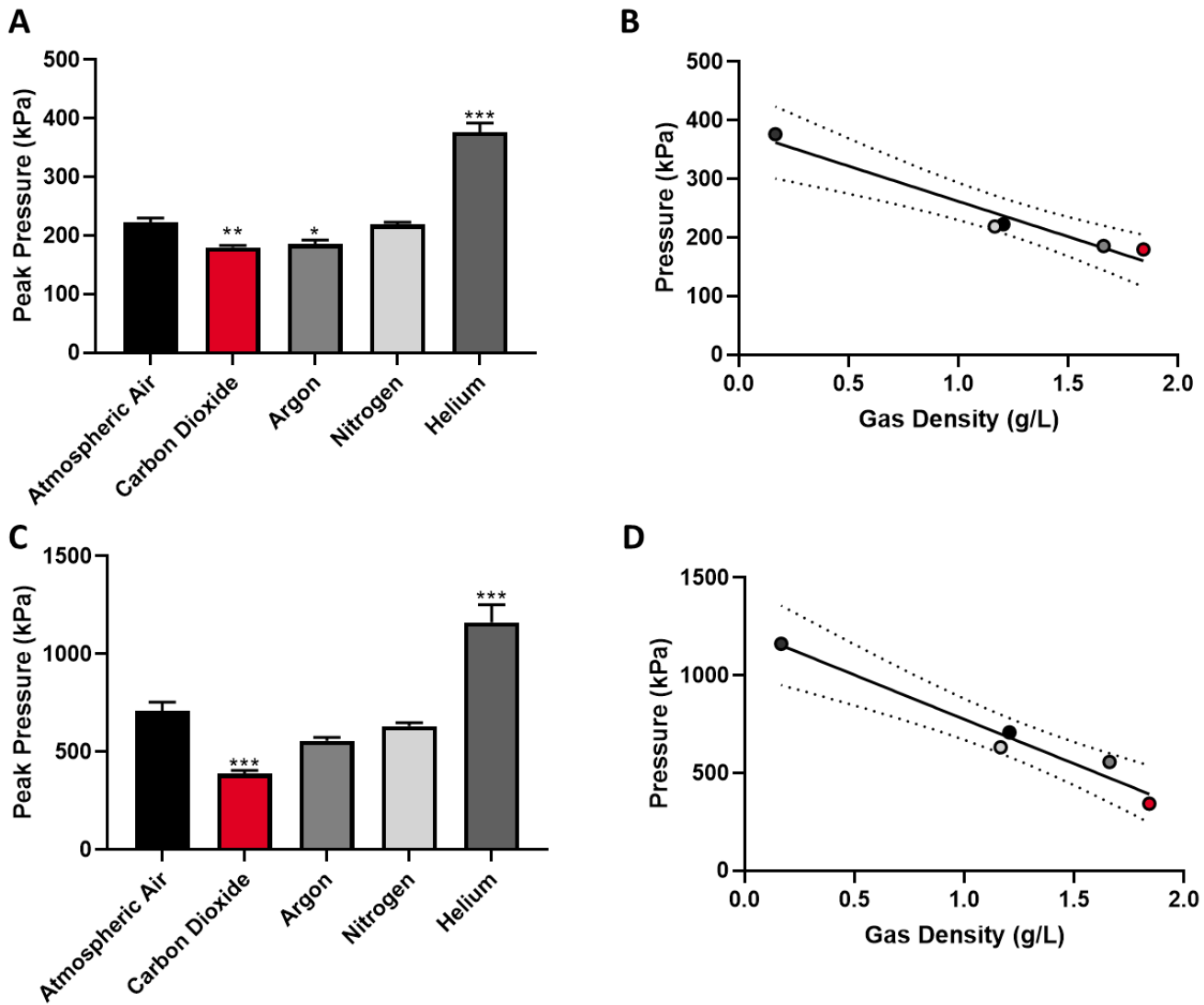

**Figure 4. A**. Peak pressure values measured at the exit of the driven section. (Ordinary one-way ANOVA, P<0.0001, followed by post hoc Dunnett's multiple comparison tests, ***P=0.0014 Atmospheric Air vs Carbon Dioxide, *P=0.0101 Atmospheric Air vs Argon, P=0.9886 Atmospheric Air vs Nitrogen, ***P<0.0001 Atmospheric Air vs Helium) **B.** Correlation between peak pressure at the exit of the driven section and the density of each gas (Linear regression, $R^2$=0.96) **C.** Peak Incident pressures for each gas used measured 5 cm from the exit of the shock tube. (Ordinary one-way ANOVA, P<0.0001, followed by post hoc Dunnett's multiple comparison tests, **P<0.0001 Atmospheric Air vs Carbon Dioxide, P=0.0668 Atmospheric Air vs Argon, P=0.4972 Atmospheric Air vs Nitrogen, ***P<0.0001 Atmospheric Air vs Helium) **D.** Correlation between peak incident pressure and the density of each gas (Linear regression, $R^2$=0.97).

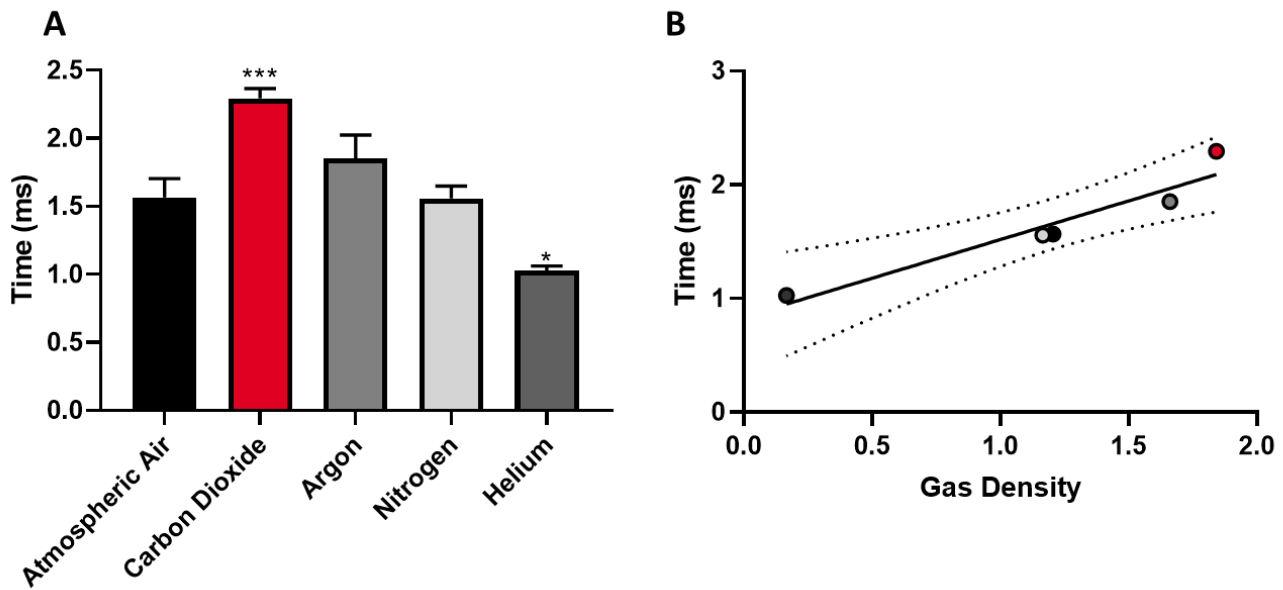

**Figure 5.** The duration of the positive phase is dependent on the density of the driver gas used. **A**. Duration of the positive phase (Ordinary one-way ANOVA, P<0.0001, followed by post hoc Dunnett's multiple comparison tests, ***P=0.005 Atmospheric Air vs Carbon Dioxide, P=0.2803 Atmospheric Air vs Argon, P>0.9999 Atmospheric Air vs Nitrogen, *P=0.0168 Atmospheric Air vs Helium). **B**. Correlation between positive phase duration and gas density (Linear regression, $R^2$=0.91)

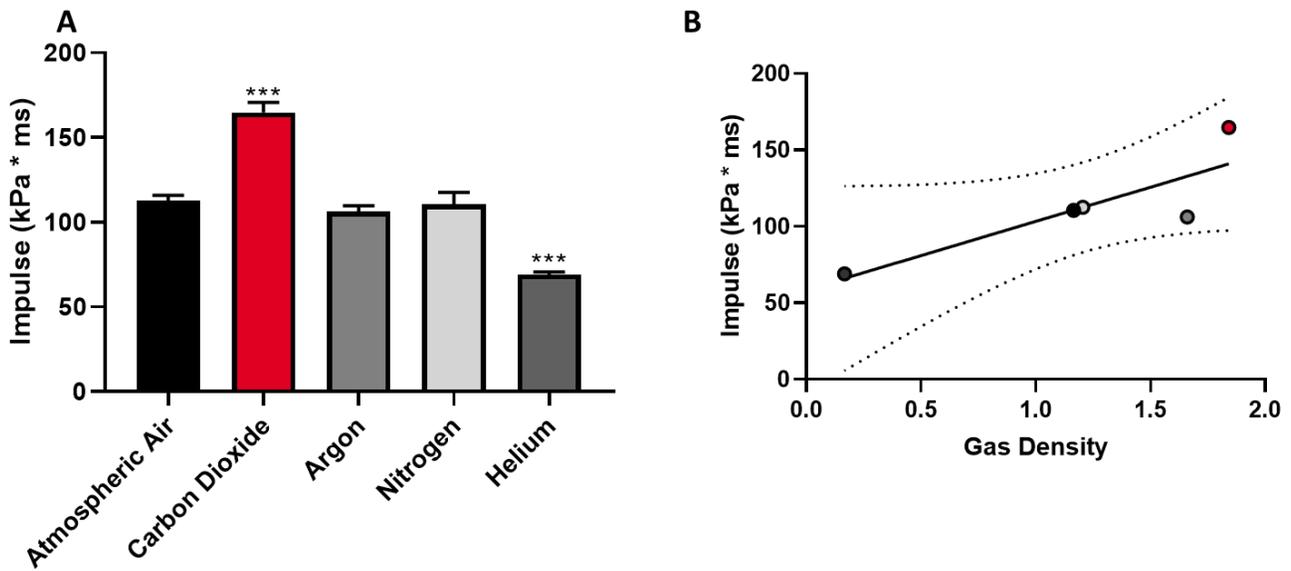

**Figure 6**. The positive impulse of the blast wave is weakly correlated with the density of the driver gas. **A**. Positive impulse calculated for each of the five gases (Ordinary one-way ANOVA, P<0.0001, followed by post hoc Dunnett's multiple comparison tests, ***P<0.0001 Atmospheric Air vs Carbon Dioxide, P=0.8512 Atmospheric Air vs Argon, P=0.9961 Atmospheric Air vs Nitrogen, ***P=0.0001 Atmospheric Air vs Helium). **B**. Correlation between positive impulse and gas density (Linear regression, $R^2$=0.72)

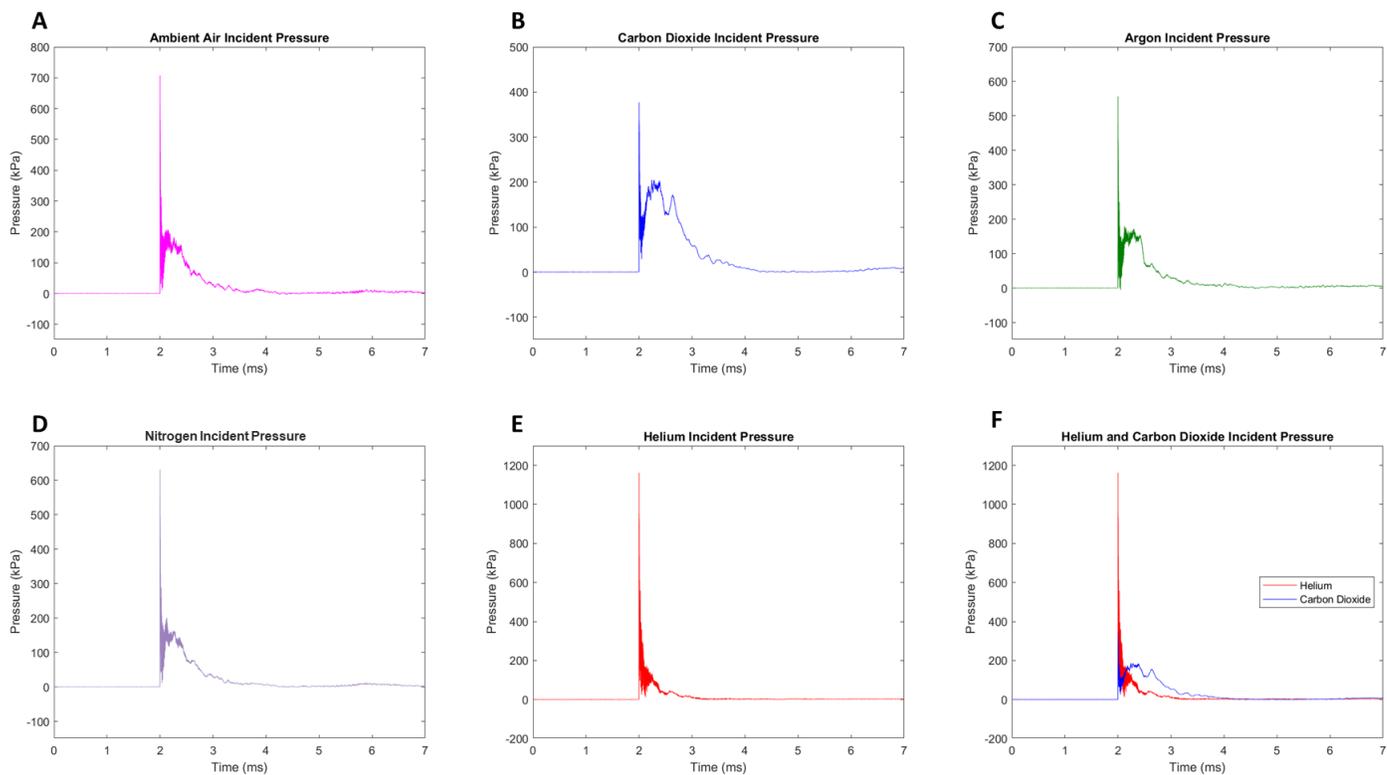

**Figure 7.** The average pressure vs time curves produced for each of the driver gases: **A**. Atmospheric air, **B**. Carbon dioxide, **C**. Argon, **D**. Nitrogen, **E**. Helium. **F**. An overlay of the curves for helium and carbon dioxide highlights the differences between a low density gas and a high density gas.

| Driver Gas | Density at NTP (g/L) |
|---|---:|
| Helium (He) | 0.1664 |
| Medical-Grade Atmospheric Air | 1.204 |
| Nitrogen ($N_2$) | 1.165 |
| Argon (Ar) | 1.661 |
| Carbon Dioxide ($CO_2$) | 1.842 |

**Table 1**. Gasses used and their corresponding densities at normal temperature and pressure (NTP; 101.325 kPa and 20°C).